\documentclass[twocolumn,showpacs,preprintnumbers,amsmath,amssymb]{revtex4}

\usepackage{epsfig}

\begin{document}


\title{A nonlinear dynamical model of human gait}
\author{Bruce J. West$^{1,2,3}$ and Nicola Scafetta.$^{1,2}$ }
\date{\today}

\affiliation{$^{1}$ Pratt School of EE Department, Duke University, P.O. Box
90291, Durham, NC 27701. } 
\affiliation{$^{2}$ Physics Department, Duke
University, Durham, NC 27701. } 
\affiliation{$^{3}$ Mathematics Division,
Army Research Office, Research Triangle Park, NC 27709. }

\begin{abstract}
We present a nonlinear stochastic model of the human gait control system in
a variety of gait regimes. The stride interval time series in normal human
gait is characterized by slightly multifractal fluctuations. The fractal
nature of the fluctuations become more pronounced under both an increase
and decrease in the average gait. Moreover, the long-range memory in these fluctuations is lost
when the gait is keyed on a metronome. The human locomotion is controlled by a
network of neurons capable of producing a correlated syncopated output.
The central nervous system is coupled to the motocontrol system, and together they control the
locomotion of the gait cycle itself.  The metronomic gait is simulated by a
forced nonlinear oscillator with a periodic external force associated
with the conscious act of walking in a particular way.
\end{abstract}



\maketitle

\section{Introduction}

Walking is a complex process which we have only recently begun to understand
through the application of nonlinear data processing techniques \cite{hausdorff1,hausdorff99,west1,west2} to study interval data. It has been known for over a century
that there is variation of 3-4\% in the stride interval of humans during
walking \cite{vierordt}, and only in the last decade did Hausdorff et al. 
\cite{hausdorff1} demonstrate that the stride-interval time series exhibits
long-time correlation, suggesting that the phenomenon of walking is a
self-similar, fractal activity. Subsequent studies by West and Griffin \cite
{west1,west2} supported the conclusion that the human gait time series is
fractal. However, more recently it was determined that these time series,
rather than being monofractal, are weakly multifractal \cite
{griffin,scafetta1}.

Human locomotion is known to be a voluntary process, but it is also
regulated  through a network of neurons called a
Central Pattern Generator (CPG) \cite{collins1}, capable of producing a
syncopated output. The early nonlinear dynamical models of CPGs for gait
assumed that a single nonlinear oscillator be used for each limb
participating in the locomotion process \cite{collins2}. Therefore a
quadraped required the coupling of four nonlinear oscillators to determine
the correct phase relations among the four legs in order to distinguish
between various modes of locomotion, that is, walking,  trotting, cantering  and
galloping. More recent dynamical models, using the property of
synchronization of nonlinear dynamical systems, allow for neurons within an
assembly to become enslaved to a single rhythmic muscular activity. Thus,
rather than having a separate nonlinear oscillator for each limb, it is
possible to have a single CPG to determine how we walk. 

The model that  we present here, the super CPG (SCPG), assumes that the central nervous
system is coupled to the motocontrol system, and together they control the
locomotion of the gait cycle. We stress that it is the period of the gait cycle that
is ultimately measured in these stride interval  experiments, and not the neural firing activity.
The dynamics of human gate may also be voluntarily forced, for example, by
following the frequency of a metronome. We model the complex gait system by
assuming that the amplitude of the impulses of the correlated  firing neural centers
regulate only the unperturbed inner frequency of a nonlinear forced Van der Pol
oscillator \cite{holmes} that mimics the gait cycle. The stride interval is
assumed to coincide with the actual period of the Van der Pol oscillator. In
this way the  gait frequency may differ slightly from the  potential frequency induced by the neural firing activity.  In fact, the chaotic behavior of nonlinear
oscillators, like the Van der Pol oscillator,  allows a more complex behavior that may be controlled also
by an constraint that forces the oscillator to follow a particular fixed
frequency.

The SCPG model is tested on a data set available in the
public domain archives Physionet \cite{physionet}. These data were
originally collected and used by Hausdorff et al. \cite{Hausdorff4} to
determine the dependence of the fractal dimension of the time series on
changes of the average rate of walking. These data  contain the stride
interval time series for 10 healthy young men walking at a slow, normal and
fast pace, for a period of one hour. The same individuals are then requested
to walk at a pace determined by a metronome set at the average slow, normal
and fast paces for 30 minutes to generate a second data set.

By estimating the H\"{o}lder exponents and their spectra using wavelet
transform \cite{struzik}, it was shown that the stride interval time series
is weakly multifractal with a main fractality close to that of the
1/f-noise. The time series is sometimes non-stationary and its fractal
variability changes in the different gait mode regimes \cite{scafetta1}. In
particular, the persistence, as well as the multifractality of the stride
interval time series, tends to increase for both slow and fast pace, above
that of the normal paces. Moreover, if the pace is constrained by a
metronome, the stochastic properties of the stride interval time series
change significantly, from persistent to antipersistent fluctuations, but,
in general, in each case there is a reduction in the long-term memory and an
increase in randomness.

In Sec. 2 we give a short introduction to the phenomenon of locomotion, the
traditional methods for modeling using the CPG, and review the data
processing used to establish the fractal behavior of stride  interval
time series. Sec. 3 reviews the stochastic properties of the  normal
and metronomic gait under different various pace velocities, slow, normal
and fast. In Sec. 4 we present the mathematical details of the SCPG model. In Sec. 5 we  compare the results of computation using the SCPG model with
the phenomenological data. Finally, in Sec. 6 we draw some conclusions.

\section{Central pattern generator and locomotion}

Walking consists of a sequence of steps. These steps may be partitioned into
two phases: a stance phase and a swing phase. The stance phase is initiated
when a foot strikes the ground and ends when it is lifted. The swing phase
is initiated when the foot is lifted and ends when it strikes the ground
again. The time to complete each phase varies with the stepping speed. A
stride interval is the length of time from the start of one stance phase to
the start of the next stance phase.

Traditionally the legged locomotion of animals is understood through the use
of a CPG, an intraspinal network of neurons capable of producing a
syncopated output \cite{collins1,crago}. The implicit assumption in such an
interpretation is that a given limb moves in direct proportion to the
voltage generated in a specific part of the CPG. Experiments establishing
the existence of a CPG have been done on animals with spinal cord
transections. Walking, for example, in a mesencephalic cat, a cat with its
brain stem sectioned rostral to the superior colliculus, is very close to
normal, on a flat, horizontal surface, when a section of the midbrain is
electrically stimulated. Stepping continues as long as a train of electrical
pulses is used to drive the stepping. This is not a simple linear response
process because changing the
frequency of the driver has little effect of the walking cycle \cite{mann}. However, since the frequency of the stepping increases in
proportion to the amplitude of the stimulation, we can conclude that the variation in the stride interval of humans  is related to the fluctuation of the amplitude of the impulses of the firing neural centers.

As Collins and Richmond \cite{collins1} point out, in spite of the studies
establishing the existence of a CPG in the central nervous system of
quadrupeds, such direct evidence does not exist for a vertebrate CPG for
biped locomotion. Consequently, these and other authors have turned to the
construction of models, based on the coupling of nonlinear oscillators, the {\it hard-wired} CPG, to
establish that the mathematical models are sufficiently robust to mimic the
locomotion characteristics observed in the movements of segmented bipeds 
\cite{cohen}, as well as in quadrupeds \cite{collins2}. These
characteristics, such as the switching among multiple gait patterns, is
shown to not depend on the detailed dynamics of the constituent nonlinear
oscillators, nor on their inter-oscillator coupling strengths \cite{collins1}.

As we mentioned in the Introduction, it has been known for over a century
that there is a variation in the stride interval of humans during walking of
approximately 3-4\% \cite{vierordt}. This random variability has been shown 
\cite{Hausdorff4,hausdorff1,west1,west2,griffin} to exhibit long-time
correlations, and suggested that the phenomenon of walking is a
self-similar, fractal, activity. The existence of fractal time series better
suggests that the nonlinear oscillators needed to model locomotion operates
in the unstable, that is, in the chaotic regime.

A stochastic version of a CPG was developed by Hausdorff et al. \cite
{Hausdorff4,griffin} to capture the fractal properties of the inter-stride
interval time series. This stochastic model was later extended by Ashkenazy
et al. \cite{ashkenazy} to describe the changing of gait dynamics as we
develop from childhood to adulthood. The model is essentially a random walk
on a correlated chain, where each node of the chain is a neural center of
the kind discussed above, and with a different frequency. This random walk
is found to generate a fractal process, with a multifractal width that
depends parametrically on the range of the random walker's step size. Ashkenazy et
al. \cite{ashkenazy} focused on explaining the changes in the gait time
series during maturation, using the stochastic CPG model.

Herein we extend the previous models by assuming that gait dynamics are
regulated by a stochastic correlated CPG similar to that of Ashkenazy et al. 
\cite{ashkenazy}, coupled to the nonlinear oscillators needed to model
locomotion in the unstable, forced and chaotic regimes. We show that two
parameters, the average frequency $f_{0}$ and the intensity $A$ of the
forcing component of the nonlinear oscillator, are sufficient to determine
both the fractal and multifractal variability of human gait under normal,
stressed and metronomic conditions, using the SCPG model.

\section{Human gait analysis}

In this section we summarize the main fractal and multifractal
characteristics of the stride interval of the human gait  data discussed in detail
elsewhere \cite{scafetta1}. We have analyzed the gait time series of 10
persons in the three different conditions of slow, normal and fast walking.
Each time series is approximately one hour long for unconstrained walking,
see, for example Fig. 1, for slow, fast and normal walking. Similarly, each
time series is approximately 30 minutes long for metronomically constrained
walking, see, for example, Fig. 2, for slow, fast and normal walking.
Participants in the study had no history of any neuromuscular, respiratory
or cardiovascular disorders. They were not taking any medications and had a
mean age of 21.7 years (range: 18-29 years); mean height $1.77\pm 0.08$
meters and mean weight $71.8\pm 10.7$ kg. All subjects provided informed
written consent. Subjects walked continuously on level ground around an
obstacle-free, long (either 225 or 400 meters), approximately oval path and
the stride interval was measured using ultra-thin, force sensitive switches
taped inside one shoe. For the metronomic constrained walking, the
individuals were told only once, at the beginning of their walk, to
synchronize their steps with the metronome. More details regarding the
collection of data can be found at Physionet \cite{physionet} from where the
data were downloaded and in Ref. \cite{scafetta1,Hausdorff4}.

The fractal and multifractal analysis of the data is done by studying the
estimated distribution of the local H\"{o}lder exponents using wavelet
transforms \cite{struzik,Mallat,arneodo2,Stanley1}. To better understand the
meaning of the H\"{o}lder exponent $h$ we recall that the relation between the
H\"{o}lder and Hurst exponent $H$ \cite{2Mandelbrot} in the continuum limit of a
monofratal noise is $h=H-1$ according to the notation adopted in \cite
{scafetta1}. Consequently, $h=0$ corresponds to pink or 1/f-noise; $%
-1<h<-0.5 $ corresponds to antipersistent noise; $h=-0.5$ corresponds to
uncorrelated Gaussian noise; $-0.5<h<0$ corresponds to correlated noise; $%
h=0.5$ corresponds to Brownian motion and $h=1$ corresponds to black noise.
Therefore, the fractal properties of the data can be studied by determining
the mean value of the distribution of  H\"{o}lder exponents and the multifractal
properties are given by the width of the distribution itself. However, we
stress that a time series of finite length will have a H\"{o}lder-exponent
distribution with a non-zero width. The existence of such a non-zero width
can be a source of confusion between a monofractal time series of finite
length and a truly multifractal time series. A multifractal time series can
only be distinguished from a monofractal time series of the same length, if
the width of its H\"{o}lder-exponent distribution is significantly larger
than that of a correspondent monofractal time series \cite{scafetta1}.

Typical distributions of H\"{o}lder exponents, for unconstrained walking of
a single individual, are depicted in Fig. 3. Fig. 4 shows the average
distributions of the H\"{o}lder exponents for the cohort of 10 walkers.
Figs. 3 and 4 show that stride interval time series for human gait are
characterized by strong persistent fractal properties very close to that of
1/f-noise, $h\approx 0$. However, normal gait is usually slightly less
persistent than both slow and fast gait. The slow gait has the most
persistent fluctuations and may present non-stationary properties, $h>0$.
The slow gait fluctuation may deviate most strongly from person to person.
The higher values of the H\"{o}lder exponents for both slow and fast gait,
relative to normal gait, may be explained as due to a stress condition
that increases the persistency and, therefore, the long-time correlation of
the fluctuations. Moreover, the regular curves of Fig. 4 show that
unconstrained walking is characterized by fractal properties that do not
change substantially from one individual to another. Finally, a careful
comparison of the widths of the distributions of H\"{o}lder exponents for
the different gaits with the widths for a corresponding monofractal noise of
the same length has proven that the stride interval of human gait is only
weakly multifractal \cite{scafetta1}. However, the multifractal structure is
slightly more prominent for fast and slow gait than for normal gait.

Fig. 5 shows typical distributions of the H\"{o}lder exponents for
metronome-constrained walking, which is little different from the histograms
in Fig. 3. Fig. 6 shows the average distributions of the H\"{o}lder
exponents for all 10 walkers. The figures clearly indicates that under the
constraint of a metronome, the stride interval of human gait become more
random and the strong long-time persistence of the 1/f-noise is lost for
some individuals. The data present a large variability of the H\"{o}lder
exponents from persistent to antipersistent fluctuations, that is, the
exponent spans the entire range of $-1<h<0$. However, the metronome
constraint usually has a relatively minor effect upon individuals walking
normally, the second peak at low H\"{o}lder exponents in Fig. 6 being attributable to a single person,
who has difficulty with the external cadence. Probably, by walking at a
normal speed an individual is more relaxed and he/she walks more naturally.
The fast gait appears to be almost uncorrelated noise because the distribution of
H\"{o}lder exponents is centered close to $h=-0.5$ characteristic of
Gaussian or uncorrelated random noise. Finally, the slow gait presents a
large variability from persistent to untipersistent fluctuations.

We notice that some individuals may be unable to walk at a given cadence
and their attempts to synchronize the pace results in a continual
shifting of the stride interval longer and shorter in the vicinity of an
average. For these individuals the phasing is never right and this gives
rise to a strong antipersistent signal for all three gait velocities.

In summary, the stride interval of human gait presents a complex behavior that
depends on many factors. Walking is a strongly correlated neuronal and
biomechanical phenomenon which may be strongly influenced by two different
stress mechanisms; (a) a natural stress that increases the correlation of
the nervous system that regulates the motion at the changing of the gait
regime from a normal relaxed condition, to a consciously forced slower or
faster gait regime; (b) a psychological stress due to the constraint of
following a fixed external cadence such as a metronome. The metronome breaks
the long-time correlation of the natural pace and generates a large fractal
variability of the gait regime. In the next section we present a
multifractal CPG model able to reproduce these properties.

\section{The SCPG model for human gait}

In this section we introduce a model of locomotion that governs the stride
interval time series for human gait. As anticipated in the previous sections
the model has to simulate a CPG \cite{collins1} capable of producing a
syncopated correlated output associated with a motocontrol process of the
gait cycle. Moreover, the model incorporates two separate and
distinct stress mechanisms. One stress mechanism, that has an 
\textit{internal} origin, increases the correlation of the time series due to the
change in the velocity of the gait from normal to the slower or faster
regimes. The second stress mechanism, has an \textit{external} origin, and
decreases the long-time correlation of the time series under the frequency
constraint of a metronome. We model this complex phenomenon by assuming that the
intensity of the impulses of the firing neural centers regulate only the inner
virtual frequency of a  forced Van der Pol oscillator \cite{holmes}. 
The observed stride interval is assumed to
coincide with the actual period of each cycle of the Van der Pol oscillator;
a period that depends on the unperturbed inner frequency of the oscillator, the
amplitude of the forcing function and the frequency of the forcing function.

Since the frequency of the stepping increases in
proportion to the amplitude of the electric stimulation \cite{mann}, we can assume that the time series of the intensity of the impulses fired by the neural centers is associated to a time series of a virtual frequencies$\{f_j\}$. So, in the spirit of the model suggested by Ashkenazy et al. \cite{ashkenazy},
we assume that the long-time correlated frequency of the SCPG is described
by a random walk on a finite-size correlated chain, where each node of the
chain is a neural center of the kind discussed above,  that fires an impulse with a particular intensity that is associated to a paticular virtual frequency. Ashkenazy et al. \cite{ashkenazy} focused on explaining
the multifractal changes in the gait time series during maturation from
childhood to adulthood, assuming that neural maturation is parametrically
associated with the range $C$ of the Brownian process that activates the nodes of the finite-size correlated chain of frequencies. Here, we adopt a
different approach because we are interested in modeling the gait for human
adults operating under different conditions. We assume  neural
maturation and, therefore, the standard deviation $C$ of the random walk process remains constant, whereas
the strength of the correlation among the neural centers increases with the
change of the velocity of the gait from the normal to the slower or faster
regimes. The change of velocity is interpreted as a biological stress.
Moreover, contrary to Ashkenazy et al. \cite{ashkenazy} we do not add any
noise to the output of each node to mimic biological noise. The final output given by the actual frequencies of the gait cycle fluctuates due to the chaotic solutions of the nonlinear
oscillators in the SCPG, here that being the forced Van der Pol oscillator.
The advantage of using chaos in the model, rather than noise, is that chaos
is an intrinsic property of the SCPG dynamics and therefore introduces
variability in a controllable way.

We observe that nonlinear oscillators may present chaotic regimes and may be
forced by an external frequency \cite{holmes}, so they may be useful not
only to describe the change of phase from the walk, trot, canter and gallop of the
quadrupeds, but also the variability of the stride intervals observed in
humans. In  bipeds it is possible to mimic the movements of the two legs
with two nonlinear coupled oscillators. However, because the geometry of the
bipeds' gait, contrary to that for quadrupeds, is unique and the two legs
must be shifted by $\pi$ rad in phase, we can mimic the biped's gait with
only one nonlinear oscillator. In our model we use a well-known neuronal
oscillator model, that is, the forced Van der Pol  oscillator \cite
{collins1,holmes} that is defined by the following equation 
\begin{equation}
\ddot{x}+\mu \left( x^{2}-p^{2}\right) \dot{x}+\left( 2\pi f_{j}\right)
^{2}x=A~\sin \left( 2\pi f_{0}~t\right) ~.  \label{vandrpol}
\end{equation}
The parameter $p$ controls the amplitude of the oscillations, $\mu $
controls the degree of  non linearity of the oscillator, $f_{j}$ is the inner
virtual frequency of the oscillator during the $j-th$ cycle that is related to the intensity of the  the $j-th$ neural fired impulse, and $A$ and $f_{0}$ are respectively
the strength and the frequency of the external driver. The
frequency of the oscillator would be $f$ if $A=0$.

We notice that the non linear term, as well as the driver, induce
the oscillator to move around a limit cycle. The actual frequency of each
cycle may differ from the inner frequency $f$.  We assume that at the
conclusion of each cycle, a new cycle is initiated  with a new inner frequency 
$f_{j}$ produced by the  stochastic CPG model  while all other 
parameters are kept constant.
However, the simulated stride interval is given by the actual period of each cycle of
the Van der Pol oscillator.

We assume that the neural centers of the SCPG may fire impulses with different amplitudes that induce virtual
frequencies $\{f_{i}\}$ with  finite-size correlations. Here, therefore,  we model directly the time series of  virtual frequencies. The frequencies 
$\{f_{i}\}$ are centered around the driver frequency $f_{0}$ according to
the relation 
\begin{equation}
f_{i}=f_{0}+\gamma ~X_{i}  \label{fress}
\end{equation}
where $\gamma$ is a constant and $X_{i}$ is a finite-size correlated variable,
that is, 
\begin{equation}
\frac{<X_{i}~X_{i+r}>}{<X_{i}^{2}>}=\exp \left[ -\frac{r}{r_{0}}\right] ~.
\label{corrf}
\end{equation}
The parameter $r_{0}$ measures the spatial range of the correlations of the
neural network. The chain of  frequencies $f_{i}=f_{0}+\gamma~X_{i}$ is generated
by a first-order autoregressive process, also known as a linear Markov
process \cite{priestley}, that is generated by the recursion equation 
\begin{equation}
X_{i}=a~X_{i-1}+\varepsilon _{i}~,  \label{ar1procc}
\end{equation}
where $0<a<1$ is a constant and $\{\varepsilon _{i}\}$ is a normalized
zero-centered discrete Gaussian process. It is easy to prove \cite{priestley} that the
autocorrelation function of the chain $\{X_{i}\}$ is given by 
\begin{equation}
\rho (r)=\frac{<X_{i}~X_{i+r}>}{<X_{i}^{2}>}=a^{r}~.  \label{corrf2}
\end{equation}
A direct comparison between Eqs. (\ref{corrf}) and (\ref{corrf2}) gives $%
a=\exp [-1/{r_{0}}]$, so, we can easily generate a data sequence with the
desired finite-size correlation value $r_{0}$. Following Ref. \cite
{ashkenazy}, we assume that a frequency is activated by the position of a
random walker given by the function $g(j)$ with $j=1,2,\dots $, whose jump
sizes follow a Gaussian distribution of width $C.$ The width of this
distribution, according to the interpretation of Ashkenazy et al. \cite
{ashkenazy}, is associated with the human neural age maturation. This random
walk mechanism allows us to obtain from the finite-time, correlated frequency
series $\{f_{i}\},$ a new time series of frequencies $\{f_{j}\}$ with 
$i=g(j),$ characterized by long-time correlations. Finally, the sequence of frequencies $\{f_{j}\}$ is used in Eq. (\ref{vandrpol}) recursively.

Normal gait, characterized by the frequency $f_{0,n}$, is assumed to occur when the
individual is relaxed, consequently the correlations between the neuronal
centers are minimum. By implication, whether the gait increases or decreases
in velocity, the correlations between the neuronal centers increases. This
increase in the stress is modeled by  using the short-time correlation
parameter $r_{0}$ of the stochastic CPG by assuming 
\begin{equation}
r_{0}=r_{0,n}\left[ 1+B \left( f_{0}-f_{0,n}\right) ^{2}\right] ~,  \label{r0cond}
\end{equation}
where $r_{0,n}$ is the short-range correlation among the firing neural centers at the normal frequency gait and $B$ is a positive constant that measures the increasing of short-range correlation at the anomalous frequency gait.

We observe that in our experimental data set  we measure the stride
intervals of the gait. It is true that the frequency of walking may be
associated with a long-time correlated neural firing activity that induces virtual pace frequency, nevertheless
the walking is also constrained by the biomechanical motocontrol cycle that
directly controls movement and produces the pace itself. The SCPG introduced here  incorporates both the neural firing activity and the
motocontrol constraints.

In summery, the SCPG model is based on the coupling of a stochastic with
a hard-wired CPG model and depends on many factors. The most important
parameters are the short-correlation size $r_{0}$ of Eq. (\ref{corrf}), that
measures the correlation between the neuron centers of the stochastic CPG,
the intensity $A$ of the forcing driving component of the nonlinear
oscillator of Eq. (\ref{vandrpol}) and, of course, the average frequency 
$f_{0}$ of the actual pace that distinguish the slow, normal and fast gait
regimes. The other parameters, $\gamma$, $C$, $\mu $ and $p$ may be, to a
first-order approximation, kept fixed.

While the numerical simulations are left to the following section, we can
anticipate an interpretation of the two main parameters $r_{0}$ and $A$. In
fact, the short-correlation size $r_{0}$ may be interpreted as a parameter that
measures the natural correlation between the neural centers and  such short-time correlation increases under particular stress,
for example, when the velocity of the gait is slower or faster than the
normal relaxed situation. The intensity of the forcing driving component $A$
may be associated with the voluntary action of trying to follow a particular
cadence and is expected to increase under a metronomic constraint.

\section{Simulated stride interval gait}

In this section we present and comment on our computer simulations of the
stride interval of human gait under a variety of conditions. For simplicity,
we make use of the following values of the parameters. The frequency of 
normal gait is fixed at the experimentally determined value of 
$f_{0,n}=1/1.1~Hz$, so that the average period of the normal gait is 1.1
second; the frequency of the slow and fast gait are respectively 
$f_{0,s}=1/1.45~Hz$ and $f_{0,f}=1/0.95~Hz,$ with an average period of 1.45
and 0.95 seconds, respectively, that is similar to experimentally realized
slow and fast human gaits shown in Fig. 1. 

 Also the
hopping-range parameter is chosen equal to that for adults \cite
{ashkenazy}, that is, $C=25$ and kept constant. Moreover, we chose $r_{0,n}=25$ such that for $f_{0}=f_{0,n}$ we have $r_{0}=25,$ that coincides with the
correspondending value found in Ref. \cite{ashkenazy}. To generate an artificial
sequence with a variability compatible to that of the experimental sequence,
we chose $B=50$ in Eq. (\ref{r0cond}) and, in Eq. (\ref{fress}), $\gamma=0.02$ , that is a value compatible to the average of the standard
deviation of all the data analyzed by us \cite{scafetta1}, however, the
value of $\gamma$ may be smaller and may decrease with an increase in the frequency $f_{0}$ and/or
an increase in the intensity of the forcing amplitude $A$ of Eq. (\ref{vandrpol}). So, we
choose a frequency $f_{0}$, calculate $r_{0}$ via Eq. (\ref{r0cond}) and the
Markovian parameter $a$, then we generate a chain of frequencies $\{f_{i}\}$
via Eqs. (\ref{fress}) and (\ref{ar1procc}) and, finally, by using the
random walk process to activate a particular frequency of the short-time
correlated frequency neural chain, we obtain the time series of the
frequencies $\{f_{j}\}$ to use in the time evolution of the Van der Pol
oscillator. For simplicity, we keep constant the two parameters of the
nonlinear component of the oscillator (\ref{vandrpol}), $\mu =1$ and $p=1$.
The only parameters allowed to change in the model are the mean frequency $%
f_{0}$ that changes also the value of $r_{0}$ via Eq. (\ref{r0cond}), and
the intensity $A$ of the driver of the Van der Pol
oscillator (\ref{vandrpol}).

Fig. 7 shows the stride interval time series for slow, normal and fast
computer-simulated gaits using SCPG. For the simulation of normal gait we
use $A=1$ and for both slower and faster gait, we use $A=2$. We suppose that
the amplitude $A$ of the driver of the Van der Pol oscillator (\ref{vandrpol}%
) should be smaller for normal gait than that for either the slower or
faster gaits, because in our interpretation $A$ measures the magnitude of
the constraint to walk at a particular velocity. The amplitude $A$ is
smaller for normal gait because normal gait is the most relaxed, spontaneous
and, and consequently the most automatic of the three gaits. The figure
shows that the SCPG model is able to reproduce a realistic persistence and
volatility for the three gaits by simply changing the frequency of the gait
itself. In particular note the high volatility of the slow gait that is
remarkably similar to that seen in Fig.1.

Fig. 8 shows the stride interval time series for slow, normal and fast
metronome-triggered computer-simulated gaits. We use the same frequency
series generated by the SCPG used to produce the sequences of Fig. 7. We
only change the intensity $A$ of the driver of the Van der Pol oscillator (%
\ref{vandrpol}). We use for the normal gait $A=4$ and for both slower and
faster gait $A=8$. Again we suppose that the intensity $A$ of the driver of
the Van der Pol oscillator (\ref{vandrpol}) should be smaller than that for
both slower and faster gaits, because normal gait is the most relaxed and
spontaneous. By comparing Figs. 7 and 8 we note the increase in randomness,
the loss of persistency and the reduction in volatility; all effects that
are induced in the latter time series by increasing the value of $A$ and are found in the phenomenologic data shown in Figs. 1 and 2.

Fig. 9 shows histograms of distributions of the H\"{o}lder exponents for the
three computer-simulated gaits shown in Fig. 7. The calculation are done in
the same way of those used to produce the histograms in Fig. 3 for the
experimental data, for details see Ref. \cite{scafetta1}. The figure shows
that the SCPG model is able to generate artificial stride interval time
series with statistical properties similar to the fractal and multifractal
behavior of real data. By changing the gait mode from slow to normal, the
mean H\"{o}lder exponent $\overline{h}$ decreases. In the same way by
changing the gait mode from normal to fast, the mean H\"{o}lder exponent
again increases, just as it does for the real data. According to the SCPG
model, this increase in the scaling parameter is due to the increase of the
inner short-time correlation among the neuronal centers, modeled by Eq. (\ref
{r0cond}). Furthermore, this behavior is due to the biological stress of
consciously walking at a speed that is different from the normal spontaneous
speed. In addition the multifractality of the gait time series slightly
increases for a walking rate different from normal. Here again this
effect is observed in the real stride interval data and it is proven by a
slight increase in the width of the histograms for fast and in particular
slow gait.

Fig. 10 shows the histograms of probability density estimations of the H\"{o}%
lder exponents for the three metronome-triggered computer-simulated gaits
shown in Fig. 8. The calculated points show that the SCPG is able to
generate artificial stride interval time series that present similar fractal
and multifractal behaviors to those of real stride interval data taken under
the constraint of a metronome. By increasing the intensity $A$ of the driver
of the Van der Pol oscillator (\ref{vandrpol}) the randomness of the time
series increases and it is possible to obtain a large variety of time
series, from those having antipersistent to those with persistent fractal
properties. In the SCPG, the parameter $A$ measures the constraint of
consciousness on the gait, and therefore the value of $A$ has to increase if
the walker is asked to synchronize his/her pace with the frequency of a
metronome. The figure suggests that the SCPG model is able to explain a
number of other properties of the metronome-triggered walking.  Fig.
6 shows that the usually normal metronome-triggered gait is that with the
highest persistent fractal properties. Normal gait is also the most natural
under the constraint of the metronome and, therefore, we should expect that
the normal gait is the most automatic and the least constrained by human
consciousness. This
is the reason we have chosen $A=4$ for the normal metronome-triggered gait.
For both slower and faster metronome-triggered gaits we have chosen $A=8$ to
indicate a higher conscious stress, that constrains gait at anomalous
speeds. Moreover, by comparing Fig. 9 and Fig. 10 and considering that in
both simulations we have used the same value of the forcing parameter $A$
for both slower and faster gaits, we notice that the largest fractal shift
occurs for the slower gait. This increased shift implies that the slower gait is the more
sensitive to a voluntary constraint and, so, and so the slower mode has the
larger variability. In fact, our human experience and the superposition of
the distributions of H\"{o}lder exponents for the 10 cohorts in Fig. 6, show
a large fractal variability of the slower gait. Finally, Fig. 6 reveals that
few persons are characterized by a strong antipersistent pace when asked to
follow a metronome. According to the SCPG model , some people are not able
to find a natural synchronization and need to continuously adjust and readjust
 the speed of their pace to match the beat of the metronome. This
changing of pace implies a very strong conscious act and, therefore, a very
high value of the parameter $A$ that would produce a strong antipersistent
signal.

\section{Conclusion}

We have introduced a new kind of CPG model. One able to mimic
the complexity of the stride interval sequences of human gait under the several
conditions of slow, normal and fast regimes for both walking  freely and
keeping the beat of a metronome. The SCPG model is based on the
assumption that human locomotion is regulated by both the central nervous
system and by a motocontrol system. A
network of neurons produces a correlated syncopated output that is
correlated according to the level of physiological stress and this
network is coupled to the motocontrol process.  The
combination of systems controls locomotion and the variability of the gait cycle. It is the period of the gait cycle that is measured in the data sets
considered herein. Moreover,  walking may be conditioned
by a voluntary act as well, for example,  walking may be consciously
forced  following the frequency of a metronome.  We model the complex system generating the data by assuming that the
correlated firing activity of the neural centers regulate only the
inner frequency of a  forced Van der Pol oscillator. However, it 
is the forced Van der Pol oscillator that mimics the motocontrol 
mechanism of the gait cycle. The stride interval is the
actual period of each cycle of the forced Van der Pol oscillator. In this way the
 gait frequency is slightly different from the inner frequency induced by the neural firing activity  
whose inpulse intensity are able to generate only a potential, but not an actual frequency. The chaotic
behavior of such a nonlinear oscillator, like the Van der Pol oscillator,  and the possibility to force the frequency of the cycle with an external fixed frequency allows the SCPG
model to generate time series that present similar fractal and multifractal
properties to that of the human  physiological stride interval data in all
situations here analyzed. Moreover, by implementing the SCPG with four coupled forced Van der Pol oscillators as in Ref. \cite{collins1}, it should be possible to simulate the change of phase between various modes of quadrupeds' locomotion, that is, walking,  trotting, cantering  and galloping.

The variety of complex behaviors is regulated by
two parameters, the average frequency $f_{0}$ and the amplitude $A$ of the
driver of the Van der Pol oscillator. The frequency $f_{0}$ regulates the
speed and may be associated with a neuronal stress that increases the
correlation among the neural centers. The amplitude $A$ may be associated
with the voluntary action of trying to track a particular frequency and it
is expected to increase under a metronome constraint. Finally, Ref. \cite{2ashkenazy} reports that the stride interval time series for elderly subjects and for subjects with Huntington's diseases are more random than for young healthy subjects. According to the SCPG  model, this may be explained by a decrease of the normal short-range correlation among the neural centers that may be  associated with a nervous degeneration caused by injury, disease, or aging.  This decrease in correlation may be modeled through $r_{0,n}$ of Eq. (\ref{r0cond}). However, the decrease of correlation in the gait of those subjects may also be associated with an increase of the amplitude $A$ of the driving force of the Van der Pol oscillator, Eq. (\ref{vandrpol}). In fact, those subjects may also consciously choose to walk more carefully.

--------\newline
{\ {\large \textbf{Acknowledgment:}}}\newline
N.S. thanks the Army Research Office for support under grant DAAG5598D0002.

\newpage 
\begin{figure*}[tbp]
\epsfig{file=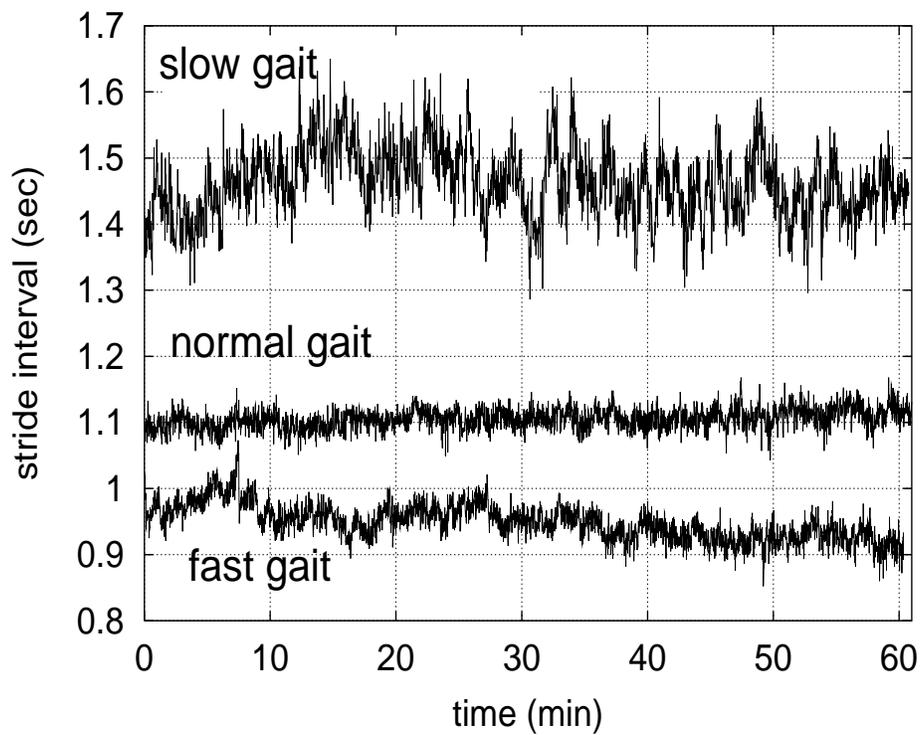,height=13cm,width=10cm,angle=-90}
\caption{ Stride interval for slow, normal and fast gait. The period of time
over which measurements were done is approximately one hour. }
\end{figure*}

\begin{figure*}[tbp]
\epsfig{file=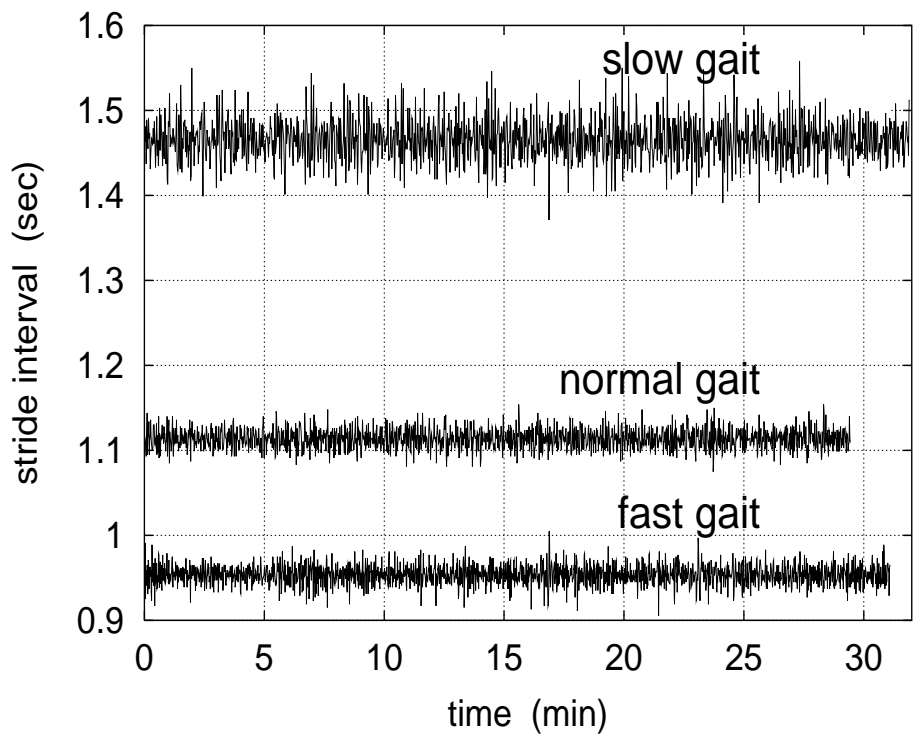,height=13cm,width=10cm,angle=-90}
\caption{ Stride intervals for slow, normal and fast gait for metronomic
triggered walking. The total period of time is approximately 30 minutes. }
\end{figure*}

\begin{figure*}[tbp]
\epsfig{file=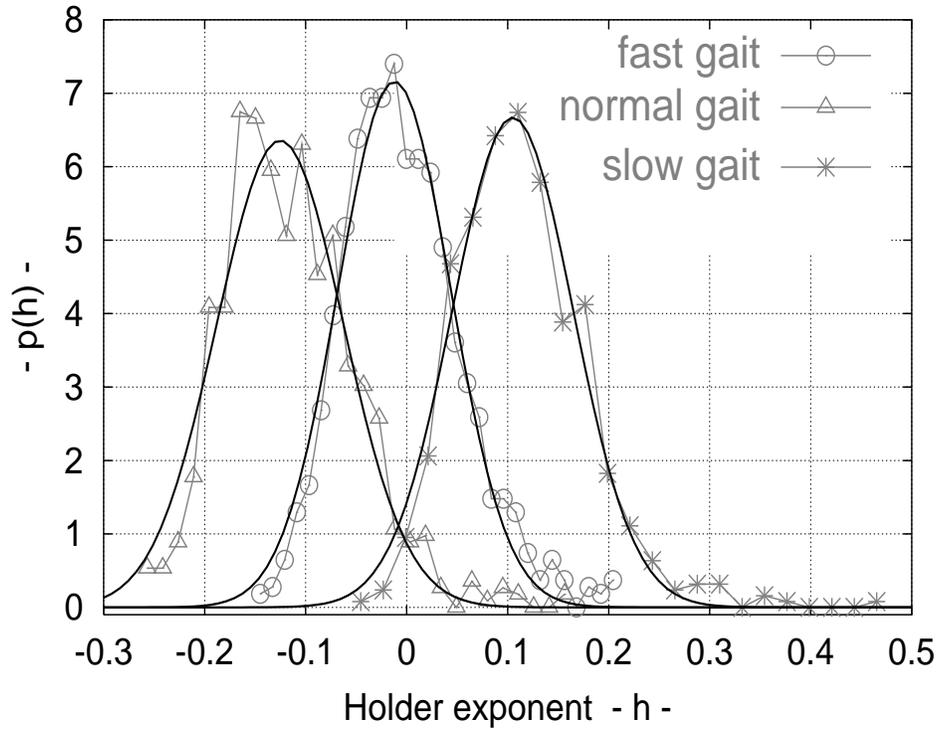,height=13cm,width=10cm,angle=-90}
\caption{ Histogram and probability density estimation of the H\"older
exponents: slow-star ($h_{0}= 0.105$, $\sigma= 0.060$), normal-triangle ($h_{0}=-0.125 $, $\sigma=0.063 $) and fast-circle ($h_{0}=-0.012 $, $\sigma= 0.056$) gait for a single individual.  The fitting curves
are Gaussian functions with average $h_{0}$ and standard deviation $\sigma$. }
\end{figure*}

\begin{figure*}[tbp]
\epsfig{file=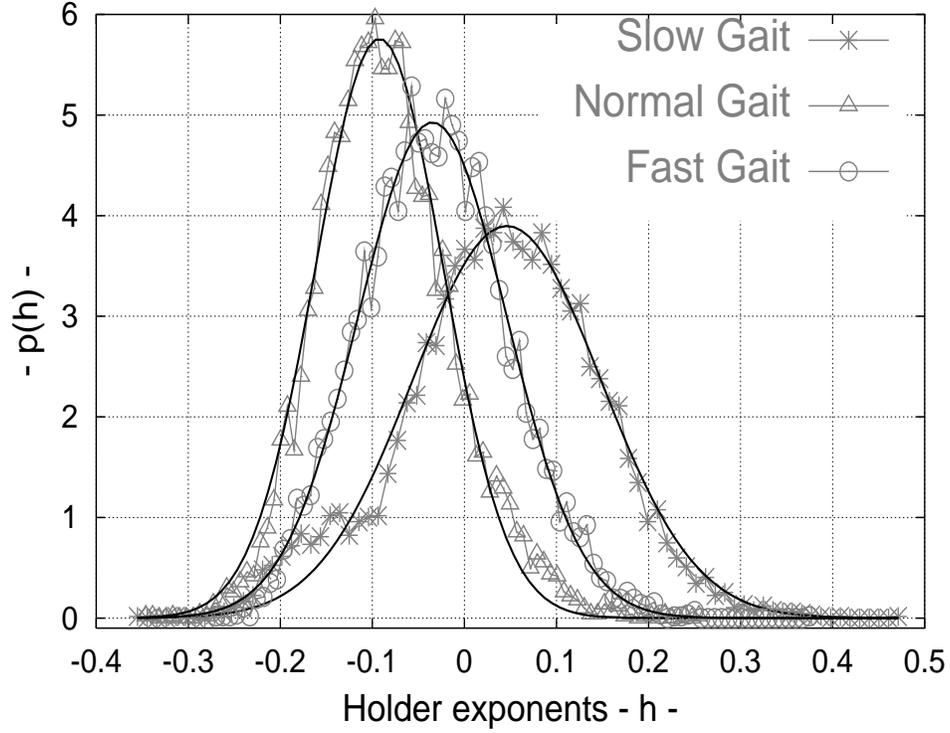,height=13cm,width=10cm,angle=-90}
\caption{Histogram and probability density estimation of the H\"older
exponents for the three walking groups are shown: slow-star, normal-triangle
and fast-circle gait. Each curve is an average over the 10 cohorts in the experiment. By changing the gate mode from slow to normal the 
Holder exponents $h$ decrease but from normal to fast they increase.   There
is also an increasing of the width of the distribution $\sigma$ by moving
from the normal to the slow or fast gait mode. The fitting curves are
Gaussian functions: slow-star ($h_{0}= 0.046$, $\sigma= 0.102$), normal-triangle ($h_{0}=-0.092 $, $\sigma=0.069 $) and fast-circle ($h_{0}=-0.035 $, $\sigma= 0.081$) gait }
\end{figure*}

\begin{figure*}[tbp]
\epsfig{file=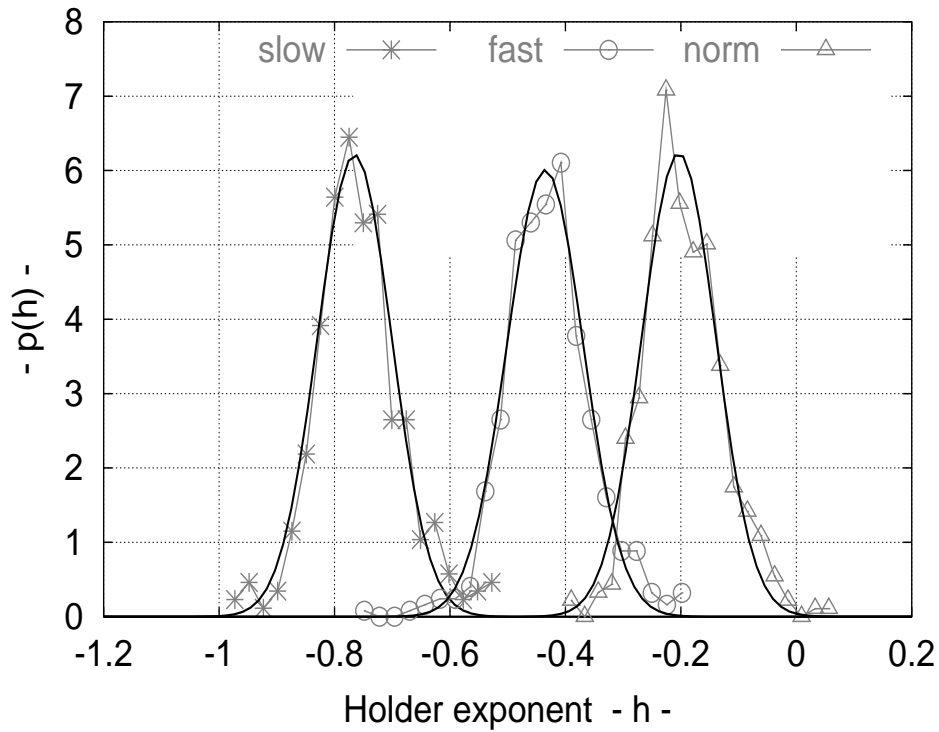,height=13cm,width=10cm,angle=-90}
\caption{ Metronomic walking for a single individual. Histogram and probability density estimation of the H\"older exponents: slow-star ($h_{0}= -0.765$, $\sigma= 0.064$), normal-triangle ($h_{0}=-0.204 $, $\sigma=0.064 $) and fast-circle ($h_{0}=-0.436 $, $\sigma= 0.066$). }
\end{figure*}

\begin{figure*}[tbp]
\epsfig{file=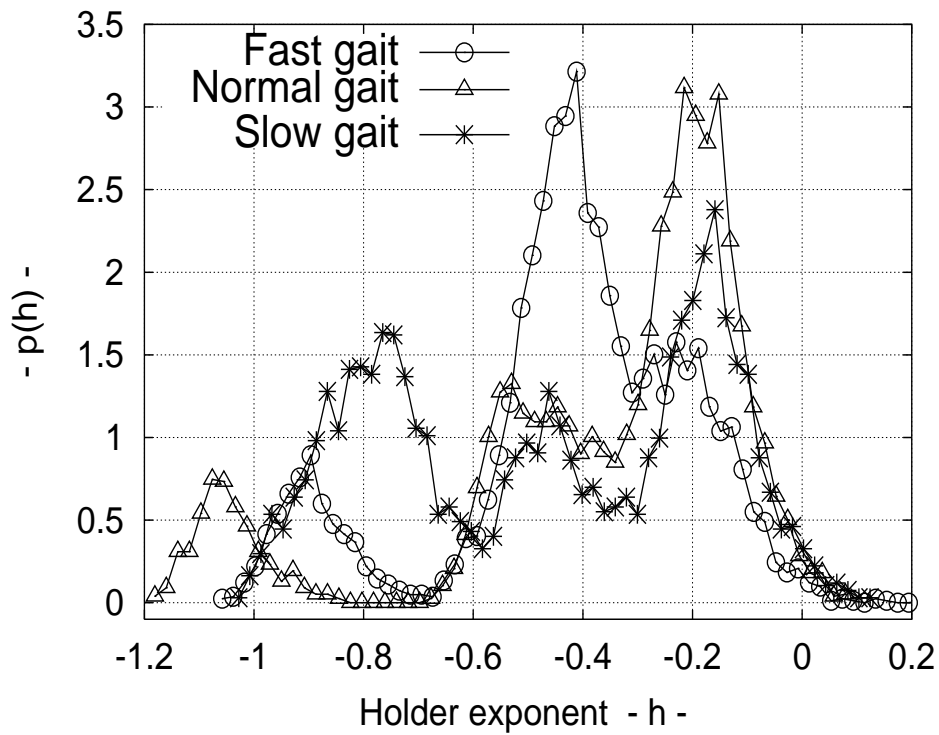,height=13cm,width=10cm,angle=-90}
\caption{Metronomic walking. Histogram estimation of the H\"older exponents
for the three walking groups: slow-star, normal-triangle and fast-circle
gait.  Each curve is an average over the 10 cohorts in the experiment. }
\end{figure*}

\begin{figure*}[tbp]
\epsfig{file=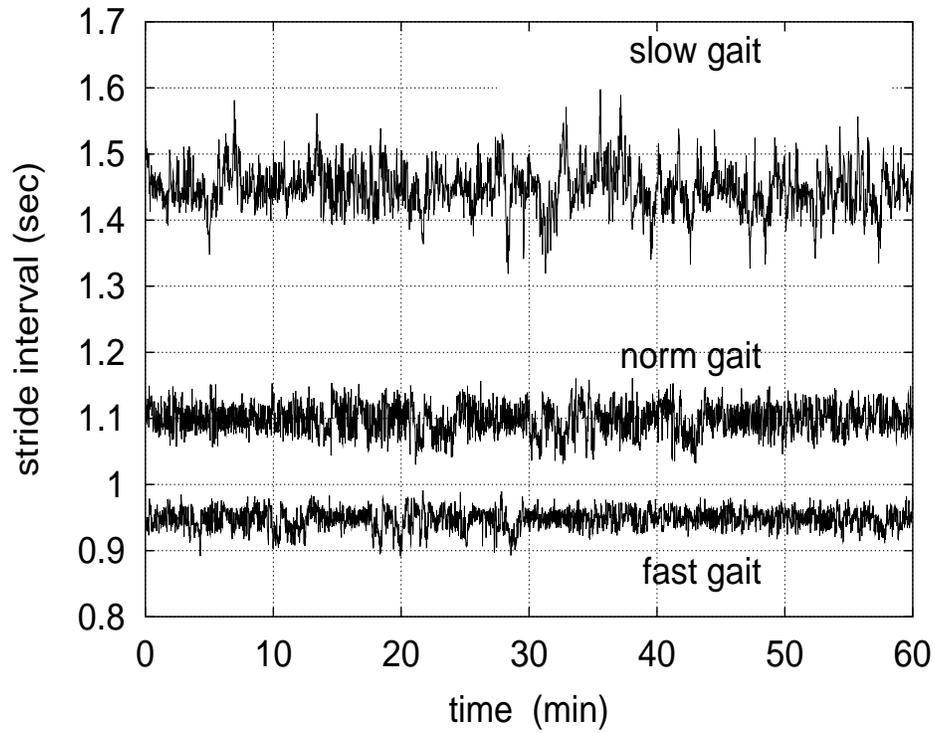,height=13cm,width=10cm,angle=-90}
\caption{ Stride interval time series for slow, normal and fast computer-simulated
gaits. }
\end{figure*}

\begin{figure*}[tbp]
\epsfig{file=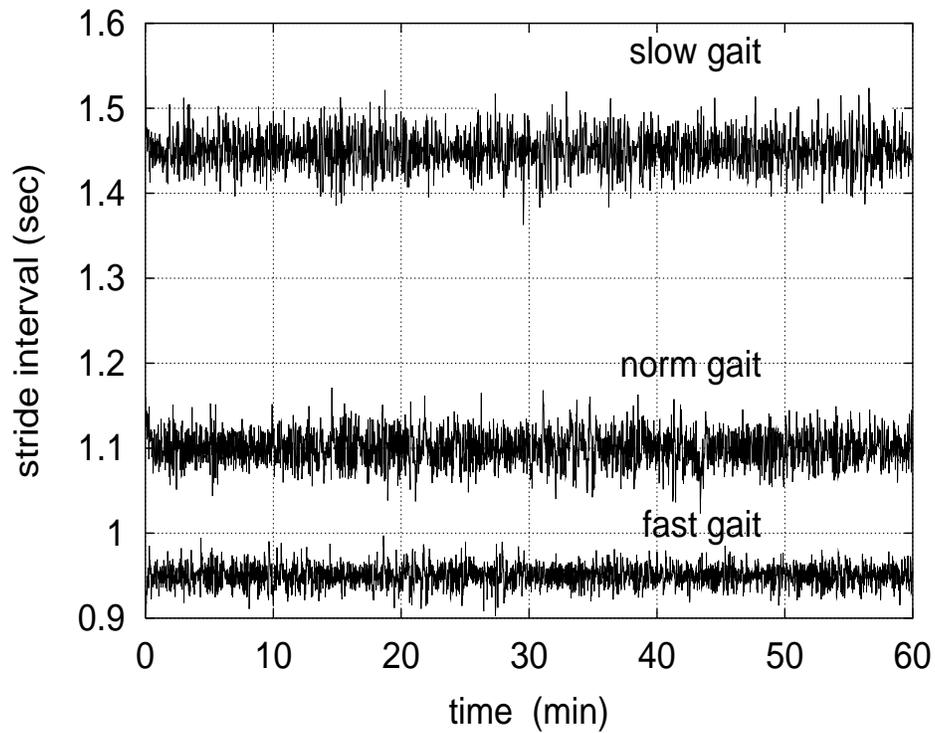,height=13cm,width=10cm,angle=-90}
\caption{ Stride interval time series  for slow, normal and fast gait for metronome-triggered computer-simulated gait. }
\end{figure*}

\begin{figure*}[tbp]
\epsfig{file=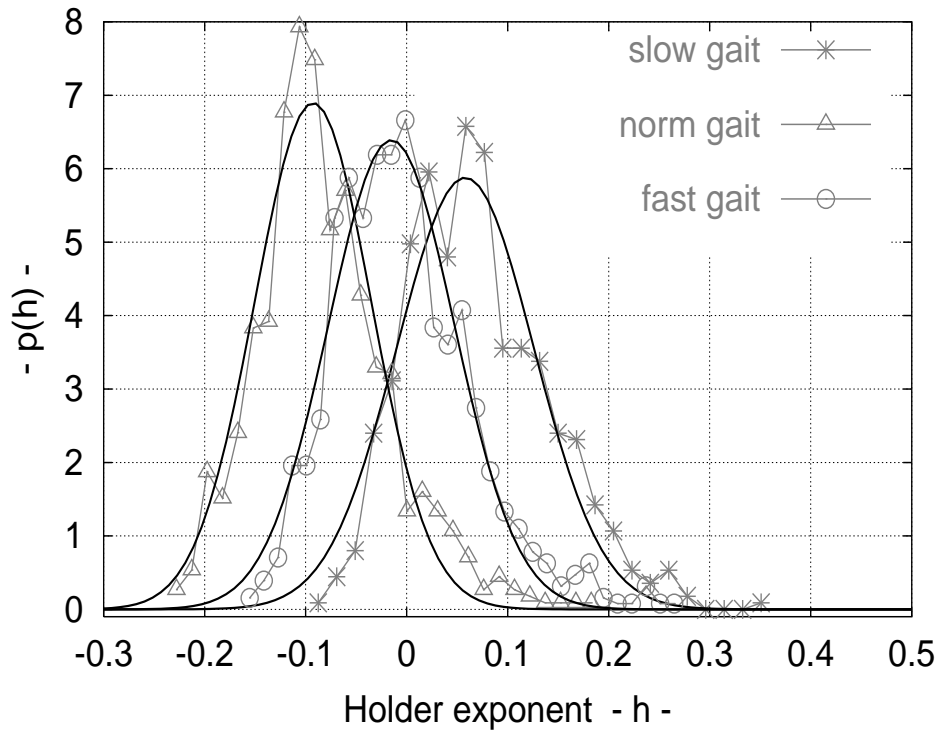,height=13cm,width=10cm,angle=-90}
\caption{ Histogram of probability density estimation of the H\"older
exponents for computer-simulated gait: slow-star ($h_{0}= 0.058$, $\sigma= 0.068$), normal-triangle ($h_{0}=-0.093 $, $\sigma=0.058 $) and fast-circle ($h_{0}=-0.015 $, $\sigma= 0.063$) gait for a single individual. }
\end{figure*}

\begin{figure*}[tbp]
\epsfig{file=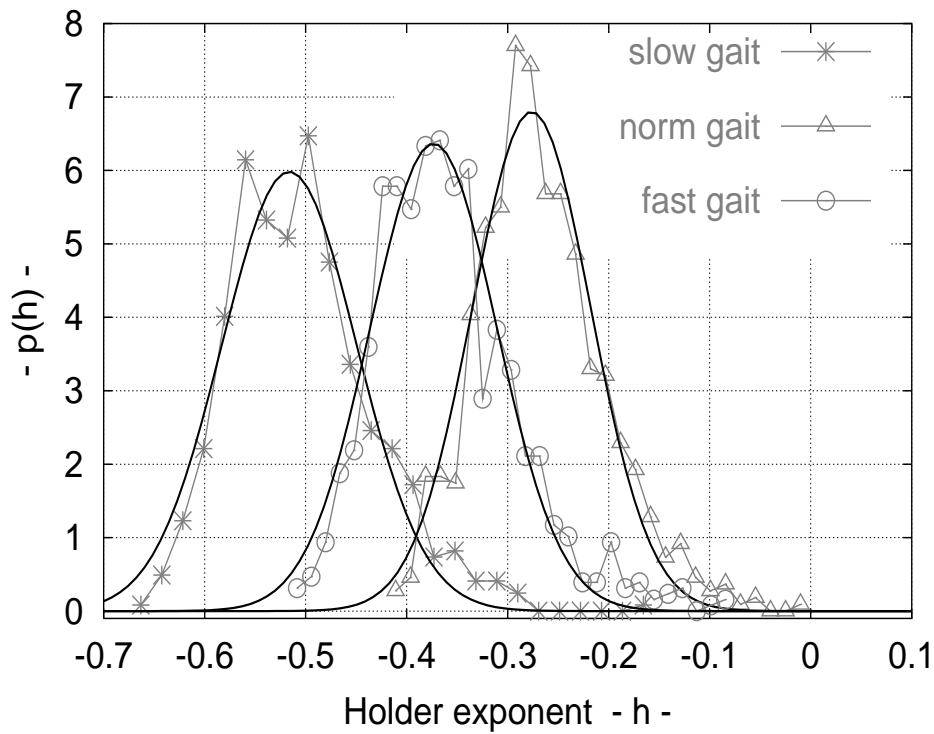,height=13cm,width=10cm,angle=-90}
\caption{ Histogram and probability density estimation of the H\"older
exponents for metronome-triggered computer-simulated gait: slow-star ($h_{0}= -0.516$, $\sigma= 0.067$), normal-triangle ($h_{0}=-0.276 $, $\sigma=0.059 $) and fast-circle ($h_{0}=-0.373 $, $\sigma= 0.063$) gait for a single individual. }
\end{figure*}

\end{document}